\documentclass[sigconf,screen,nonacm,fleqn]{acmart}
\usepackage{amsmath, nccmath}
\usepackage{mathtools}

\usepackage{soul}
\usepackage{xcolor}
\restylefloat{table}
\usepackage{tabularx}
\usepackage{listings}
\usepackage{xcolor}

\definecolor{codegreen}{rgb}{0,0.6,0}
\definecolor{codegray}{rgb}{0.5,0.5,0.5}
\definecolor{codepurple}{rgb}{0.58,0,0.82}
\definecolor{backcolour}{rgb}{0.95,0.95,0.92}
\definecolor{commentsColor}{rgb}{0.497495, 0.497587, 0.497464}
\definecolor{keywordsColor}{rgb}{0.000000, 0.000000, 0.635294}
\definecolor{stringColor}{rgb}{0.558215, 0.000000, 0.135316}
\lstdefinestyle{mystyle}{
  backgroundcolor=\color{backcolour},   
  basicstyle=\footnotesize,        
  breakatwhitespace=false,         
  breaklines=true,                 
  captionpos=b,                    
  commentstyle=\color{commentsColor}\textit,    
  deletekeywords={...},            
  escapeinside={\%*}{*)},          
  extendedchars=true,              
  frame=tb,                        
  keepspaces=true,                 
  keywordstyle=\color{keywordsColor}\bfseries,       
  language=Python,                 
  otherkeywords={*,...},           
  numbers=left,                    
  numbersep=5pt,                   
  numberstyle=\tiny\color{commentsColor}, 
  rulecolor=\color{black},         
  showspaces=false,                
  showstringspaces=false,          
  showtabs=false,                  
  stepnumber=1,                    
  stringstyle=\color{stringColor}, 
  tabsize=2,                     
  title=\lstname,                  
  columns=fixed                    
}
\lstdefinelanguage{YARA}{
  keywords={rule, meta, strings, condition, matches, rules, externals},
  keywordstyle=\color{blue}\bfseries,
  ndkeywords={and, match, callback},
  ndkeywordstyle=\color{darkgray}\bfseries,
  identifierstyle=\color{black},
  sensitive=false,
  comment=[l]{//},
  morecomment=[s]{/*}{*/},
  commentstyle=\color{purple}\ttfamily,
  stringstyle=\color{red}\ttfamily,
  morestring=[b]',
  morestring=[b]"
}
\AtBeginDocument{%
  \providecommand\BibTeX{{%
    \normalfont B\kern-0.5em{\scshape i\kern-0.25em b}\kern-0.8em\TeX}}}

\setcopyright{acmcopyright}
\copyrightyear{2022}
\acmYear{2022}
\acmDOI{XXXXXXX.XXXXXXX}

\acmConference[]{}{Woodstock, NY}
\acmPrice{}
\acmISBN{}

\begin{document}

\title{REGARD: Rules of EngaGement for Automated cybeR Defense to aid in Intrusion Response}


\author{Damodar Panigrahi}
\affiliation{%
 \institution{Mississippi State University}
  \country{Mississippi State, MS, USA}
}\email{dp1657@msstate.edu}

\author{William Anderson}
\affiliation{%
 \institution{Mississippi State University}
  \country{Mississippi State, MS, USA}
}\email{wha41@msstate.edu}

 \author{Joshua Whitman}
\affiliation{%
 \institution{Mississippi State University}
  \country{Mississippi State, MS, USA}
}\email{jsw625@msstate.edu}

 \author{Sudip Mittal}
\affiliation{%
 \institution{Mississippi State University}
  \country{Mississippi State, MS, USA}
}\email{mittal@cse.msstate.edu}

 \author{Benjamin A Blakely}
\affiliation{%
 \institution{Argonne National Laboratory}
  \country{Ankeny, IA, USA}
}\email{bblakely@anl.gov}

\renewcommand{\shortauthors}{Panigrahi, et al.}

\begin{abstract}

Automated Intelligent Cyberdefense Agents (AICAs) that are part Intrusion Detection Systems (IDS) and part Intrusion Response Systems (IRS) are being designed to protect against sophisticated and automated cyber-attacks. An AICA based on the ideas of Self-Adaptive Autonomic Computing Systems (SA-ACS) can be considered as a managing system that protects a managed system like a personal computer, web application, critical infrastructure, etc. An AICA, specifically the IRS components, can compute a wide range of potential responses to meet its security goals and objectives, such as taking actions to prevent the attack from completing, restoring the system to comply with the organizational security policy, containing or confining an attack, attack eradication, deploying forensics measures to enable future attack analysis, counterattack, and so on. To restrict its activities in order to minimize collateral/organizational damage, such an automated system must have set Rules of Engagement (RoE). Automated systems must determine which operations can be completely automated (and when), which actions require human operator confirmation, and which actions must never be undertaken. In this paper, to enable this control functionality over an IRS, we create Rules of EngaGement for Automated cybeR Defense (REGARD) system which holds a set of Rules of Engagement (RoE) to protect the managed system according to the instructions provided by the human operator. These rules help limit the action of the IRS on the managed system in compliance with the recommendations of the domain expert. We provide details of execution, management, operation, and conflict resolution for Rules of
Engagement (RoE) to constrain the actions of an automated IRS. We also describe REGARD system implementation, security case studies for cyber defense, and RoE demonstrations. 

\end{abstract}



\maketitle

\section{Introduction}

Cyber attacks are becoming increasingly sophisticated and automated. Botnets and ransomware are some of the early applications we've seen where attackers use automation as an X-factor in their campaigns. With the advent of increasingly sophisticated Artificial Intelligence (AI) /Machine Learning (ML) algorithms, we can expect that attacks will continue to mature until they are mere "button presses" for attackers. Human defenders are challenged to keep pace with even the attacks of today, let alone attacks that are dynamic, intelligent, and responsive to the actions of a defender. 

For this reason, \textit{Automated Intelligent Cyberdefense Agents (AICAs)} are an important technology to develop to combat these risks. AICA systems can be considered a type of Self-Adaptive Autonomic Computing System (SA-ACS) containing two parts: An Intrusion Detection System (IDS) and an Intrusion Response System (IRS). IDSs are widely used to detect threats to computer systems and are fundamental to identifying ongoing threats. A considerable amount of work has been done to develop and deploy IDSs. An IRS on the other hand is designed to identify a proper response to an ongoing attack automatically. 
This system deals with the problem once an intrusion has been detected. Here the goal is to identify strategies and compute a response to the questions like - \textit{How can the system be protected? Can the attack be handled in such a way that the damage is minimized?} 

To meet its goals and objectives, an IRS can compute a wide \textit{variety of potential responses} including taking actions to prevent the attack from completing, restoring the system to comply with the organizational security policy, containing or confining an attack, attack eradication, deploy forensics measures to enable future attack analysis, counterattack, etc. However, such an automated system must have defined \textit{Rules of Engagement (RoE)} to constrain its actions. Automated systems must define which actions they can take in a fully automated manner (and when), which actions require a confirmation from a human operator, and which actions must never be executed. 

In general, warfare RoEs are military directives meant to describe the circumstances under which ground, naval, and air forces will enter into and continue combat with opposing forces. Two such well-established RoE directives include the United Nations Rules of Engagement \cite{UNROE}, and the United States Department of Defense's rules of engagement (both standing peacetime ROE (SROE), and wartime ROE (WROE)). RoEs in cyberspace and cyber warfare are being debated with possible future resolutions \cite{doi:10.1080/23742917.2020.1798155}, In 2021, the United States government expressed a desire to codify an international cyberwar RoEs directive to protect critical services like telecommunications, healthcare, food and energy from cyber attacks \cite{biden, bidenputin}.




In standard kinetic doctrine, the authority to engage combatants is given to a commander. In general, action must be taken to circumvent violence, and all other options must be exhausted before offensive action may be taken. If a computer is to execute an action that will affect a computer on a foreign network, the rules need to be explicitly defined and closely examined before deployment into an active network is permissible.

Taking counter-actions requires decisions considering many factors, which is why these decisions cannot be clearly outlined for humans to follow. Judgment is required as some actions can be seen as more invasive or provocative than others, or conversely, attackers may want a particular action to be taken and they are testing the capabilities of a system. An autonomous computing agent will be capable of considering all factors and deciding a response. Still, it will be necessary for the system administrators to make clear what is \textit{not allowable} so as not to cause provocation or violate international law.


Developing a system that allows an organization to define and execute these RoEs is the main goal of this paper. We have created and implemented a rules management system called - \textit{REGARD (Rules of EngaGement for Automated cybeR Defense)} which can be integrated with an automated IRS 
and makes such agents safe to deploy in operational environments (See sections ~\ref{autonomic-computing}, ~\ref{aica-architure}, and ~\ref{irs} for more background and related works on autonomic computing, AICA, and IRS). 

\noindent \textbf{Main contribution} of this paper are:
\begin{itemize}
\item  Rules of EngaGement for Automated cybeR Defense (REGARD) system architecture, design, roles and
responsibilities, and rule definitions. The REGARD system extends the Self-Adaptive Autonomic Computing System (SA-ACS) paradigm and is based on the Automated Intelligent Cyber defense Agents (AICA) architecture. 
    
\item REGARD holds a set of Rules of Engagement (RoE) to protect the managed system according to the instructions provided by the domain expert. These rules help limit the action of the IRS on the managed system in compliance with the recommendations of the domain expert.

\item We provide details of execution, management, operation, and conflict resolution for Rules of
Engagement (RoE) to constrain the actions of an automated IRS.

\item We describe REGARD system implementation, security case studies for cyber defense, and RoE demonstrations. We showcase the execution of REGARD on two managed systems, namely, a web system and a network infrastructure management system. 
\end{itemize}

The rest of the paper is organized as follows - Section~\ref{relwork} presents background information and related work. Next, we present the REGARD system architecture and its components in Section~\ref{regard-system}. We then provide implementation details, case studies for cyber defense, and RoE demonstrations for the REGARD system in Section~\ref{regard-implementation}. Finally, we conclude the paper in Section ~\ref{conclusion}.

\section{Related Work \& Background}
\label{relwork}
This section summarizes background concepts and some related work in autonomic computing, self-adaptive systems, and Intrusion Response Systems (IRS). We also provide an overview of a generic Autonomous Intelligent Cyberdefense Agent (AICA) system architecture that influences our REGARD system (See Section \ref{regard-system}).

\subsection{Autonomic Computing System (ACS)}
\label{autonomic-computing}
An \textit{Autonomic Computing System} (ACS) is a self-managing system capable of adapting to a dynamic environment. Such systems need to meet user-defined high-level objectives \cite{1160055}. These objectives are the primary mechanism by which a user defines how a particular system operates. As a result of this paradigm, the user is free to focus on the high-level behavior of the system, such as Service Level Objectives (SLOs), while the user is free from the low-level specifics of how the system is to be implemented or optimized. An ACS should self-govern and exhibit specific characteristics described in more detail in Section ~~\ref{self-adaptive-acs}.

\begin{figure}[ht]
    \centering
    \includegraphics[scale=0.25]{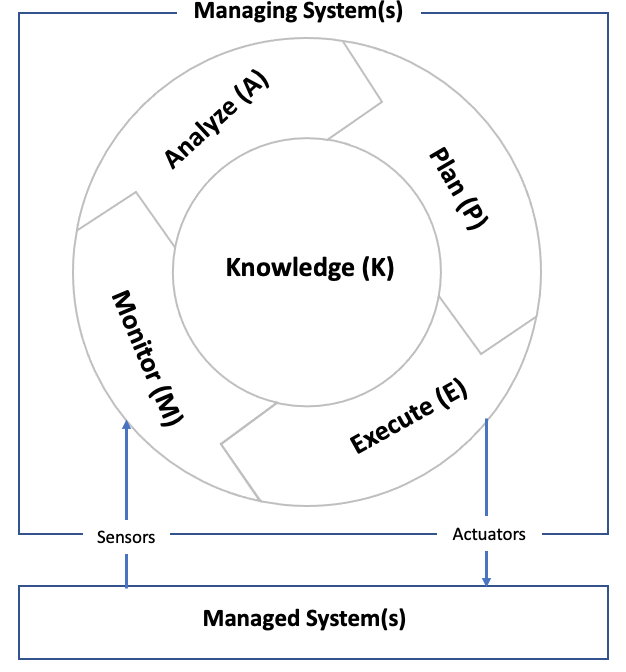}
    \caption{Autonomic computing MAPE-K feedback loop:  Autonomic computing framework's MAPE-K consists of five components, namely, Monitor (M), Analyze (A), Plan (P), and Execute (E). Managing systems collectively follows the framework. They consist of cyber security software tools, optionally utilize machine learning techniques, and work harmoniously to detect and thwart security incidents in the managed systems that provide business functionalities. }
    \label{fig:mapek}
\end{figure}

The autonomic computing paradigm in the security domain focuses on two parts: a \textit{managed system} and a \textit{managing system}~\cite{lemos2013software}. A managed system is a computer system that provides business functionality and contains application logic. A managing system is a system that is responsible for detecting and mitigating a security issue on the managed system. The managing system architecture is typically comprised of five components: Monitor, Analyze, Plan, Execute, and Knowledge. Together these components form a feedback loop called MAPE-K as shown in Figure ~\ref{fig:mapek}. The `\textit{Monitor}' component continuously surveys the managed system and stores the gathered data. Some examples of this include applications such as Prometheus~\cite{prometheus}, and Jaeger~\cite{jaeger}. The `\textit{Analyze}' component processes the gathered data and ensures compliance with the given SLOs. Examples of existing components include novelty/outlier detection implementations in ~\cite{scikit-anomaly} and STUMPY~\cite{stumpy}. The `\textit{Plan}' component brings the managed system to the desired configuration by actualizing a workflow from the current state to the desired state. The `\textit{Execute}' component is responsible for realizing the plan and performing the actions on the managed system. CFEngine~\cite{cfengine} and Ansible~\cite{ansible} are some examples of the execute component. Finally, the `\textit{Knowledge}' component stores and suitably represents the information that the other components utilize.

\begin{figure*}[ht]
    \centering
    \includegraphics[scale=0.35]{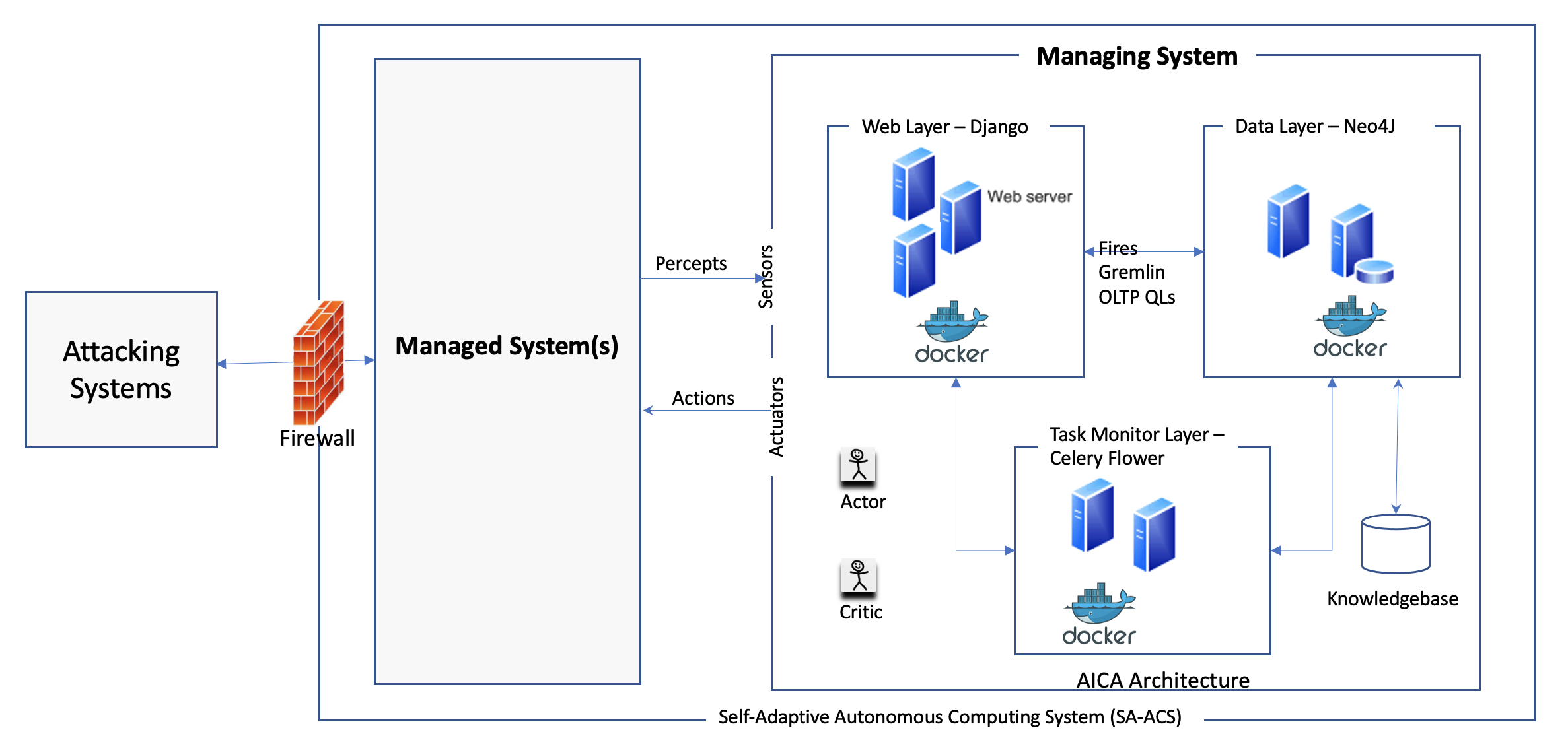}
    \caption{Automated Intelligent Cyberdefense Agents (AICA) architecture: A 3-layered AICA microservice architecture that adheres to the self-adaptive autonomous computing system specification. These layers house the MAPE-K components deployed as Docker containers. For example, the Monitor (M) components receive logs, such as Flowlogs, via the sensors.}
    \label{fig:aica-architecture}
\end{figure*}

\subsection{Self-Adaptive ACS (SA-ACS)}
\label{self-adaptive-acs}

The seminal autonomic computing system paper also introduced four distinct characteristics: self-configuration, self-optimization, self-healing, and self-protection \cite{1160055}. Recent work has extended the original set of characteristics to include features such as self-learning, self-regulation, and self-organization. \cite{poslad2009autonomous, nami2007survey}. One notable extension is the `\textit {self-adaptive}' characteristic that requires the system to adjust its behavior in response to the environment \cite{macias2013self, krupitzer2015survey}. Such an ACS is called a Self-Adaptive Autonomic Computing System (SA-ACS).

\subsection{Autonomous Intelligent Cyberdefense Agents (AICA)}
\label{aica-architure}

There are many frameworks that materialize SA-ACS \cite{cheng2009evaluating, de2015corporate, romero2013framework, basile2016complex, kott2018autonomous}. One such framework outlined in Theron et al. \cite{jajodia2020adaptive} is the Autonomous Intelligent Cyberdefense Agent (AICA). Our implementation of the AICA framework functions as a managing system (GitHub code \cite{aica-agent}). In this way, together with a managed system, they form a SA-ACS. Our managing system precepts the managed system through \textit{sensors} and acts upon the managed system through \textit{actuators}. The sensors and actuators are the AICA framework peripherals that coordinate with the Monitor and Execute components, respectively. Our AICA architecture follows the micro-service architecture pattern \cite{newman2021building}, where the application logic of each service is encapsulated in a Docker container \cite{merkel2014docker}. The AICA application can be accessed via a front-end web interface, deployed using the Django web framework \cite{django}. Our our application stores the SLOs (rules) in a Neo4J graph database \cite{neo4j}. In addition, we use Celery Flower to monitor the task execution status. Our application follows a layered architecture pattern \cite{richards2015software} with three layers, namely, a web layer, a data layer, and a monitor layer supported by Django, Neo4J, and Celery, respectively. In summary, AICA is a system for intelligently monitoring the managed system and mitigating proactively observed security threats. 

\begin{figure*}[ht]
    \centering
    \includegraphics[scale=0.35]{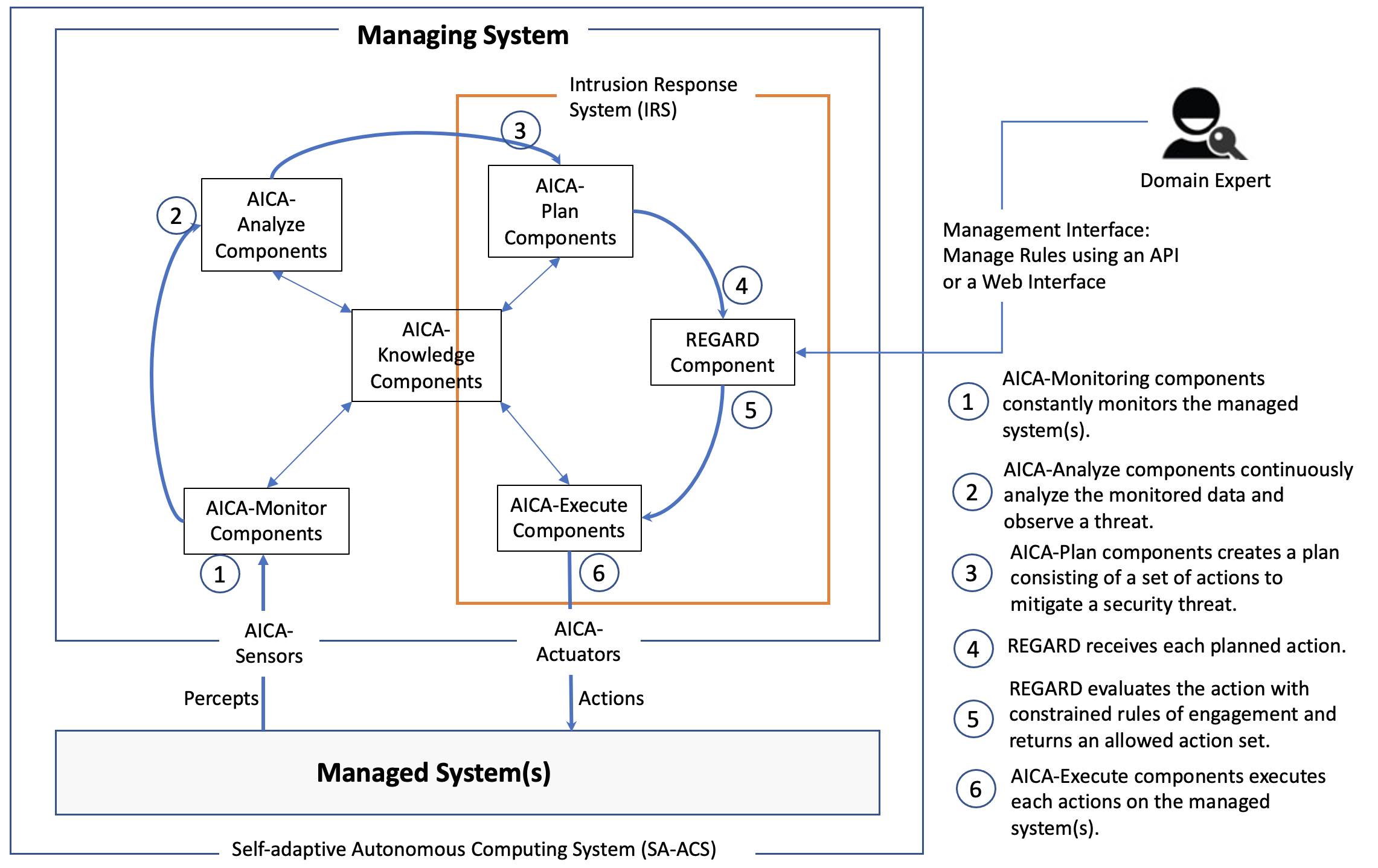}
    \caption{Rules of Engagement for Automated Cyber Defense Agents (REGARD) Component architecture: An extended AICA architecture introduces the REGARD Component. The component is an extension of the IRS system comprising the AICA-Plan and the AICA-Execute components. The core component is a Rules of Engagement (RoE) engine that enables organizational domain experts to configure descriptive rules to further allow or deny a predicted action by the AICA-Plan components.}
    \label{fig:regard-architecture}
\end{figure*}

Finally, our AICA implementation, together with a complete component-level architecture is available in a GitHub repository \cite{aica-agent}. However, a simplified high-level AICA architecture is shown in Figure ~~\ref{fig:aica-architecture}, as well as an AICA architecture fused with the MAPE-K components, henceforth referred to as REGARD, in Figure ~~\ref{fig:regard-architecture}. The SA-ACS system information flow based on the AICA architecture with REGARD integration has been described in detail in Section \ref{regard-system}.

\subsection{Intrusion Response Systems (IRS)}
\label{irs}

In the cybersecurity domain, there are two macro-components: an Intrusion Detection System (IDS) and an Intrusion Response System (IRS). Typically, an IDS detects a security threat, and an IRS mitigates it. Contemporary IDS are based on decades of research and are considered mature. Thus, many open-source, research, or commercially available IDS are available \cite{snort-ips, suricata, haas2020zeek, solarwinds}. They can, to some extent, also mitigate detected threats. However, these systems provide limited mitigation capabilities primarily because of a lack of generalizability \cite{iannucci2020hybrid,cardellini2022intrusion}. 

On the other hand an IRS aims to automatically mitigate threats and developing one is currently an active research area \cite{9180198,cardellini2022intrusion}. Here the goal is to \textit{mitigate} a threat detected by an IDS and \textit{recover} a managed system if it deviates from the desired security configuration. While considering the MAPE-K components loop for AICA, an IRS operates in the AICA-Plan and AICA-Execute components, and an IDS operates in the AICA-Monitor and AICA-Analysis components as shown in the Fig.~~\ref{fig:regard-architecture}. 

Even though IRS is an active area of research, we couldn't find any that provides a validation and constraint mechanism to planned actions by an IRS agent to boost organizational defensive mechanisms further. A few examples of the active IRS research areas are: considering non-deterministic Markov Decision Process (MDP) \cite{stefanova2018off}, improving the learning speed while considering attacker behavior \cite{paul2019learning}, considering players with partial information \cite{huang2019game}, considering the time interval between the deployment and execution of actions \cite{khoury2020hybrid}, improving the learning speed by partitioning the managed system \cite{cardellini2022intrusion}, and others \cite{bashendy2022intrusion}. The research areas are based on various IRS challenges, such as: reducing the time to act \cite{toyin2012cost}, reducing wrong actions thereby improving the managed system availability to serve business functionality \cite{shameli2014taxonomy}, providing real-time responses, reducing false-alarms, etc. \cite{inayat2016intrusion}. However, none of the challenges and research areas addresses a second layer of security: to constrain the IRS planned action.

Our work strengthens the organization's managed system security infrastructure by limiting certain IRS-planned actions which otherwise might have resulted in massive damage to the managed systems. Hence, we propose an IRS component extension to accommodate organizational admin-configured rules to overwrite an IRS agent planned action to mitigate a cyber security threat. Thus we introduce a Rules of Engagement component as depicted in Section \ref{regard-system}.

\subsection{Cybersecurity and Rule-Based Systems}
As discussed in Section ~\ref{irs}, IDSs are considered a mature, well-developed research area. Typically, IDS exist in one of two forms: anomaly-based or signature-based. Signature-based IDS leverage a set of rules to create a well-defined behavior pattern, giving rise to the authenticity and integrity of the alerts they issue. However, irrespective of the form type, they are represented in a declarative textual format as rules. These rules can follow the syntax specified by a rules engine that evaluates these rules. A few such rules syntax tools are mentioned below.

SNORT \cite{snort-ips} is both an IDS and IPS that leverages many detection technologies, all of which are supported by a rule-based language. Rules are constructed into two logical parts: actions and options. The rule action defines the set of actions to take, such as pass, log, or alert. The rule actions also define the "who, where, and what" of a packet. Once an action has been determined for a given event trigger, rule options define precisely how that action is carried out, supporting options such as messaging, logging, payload inspection, and many others. 

Another rule-based approach popular for detecting malware is YARA \cite{yara}. Condition, strings, and meta define YARA rules. The `condition component defines what conditions are to be met in order for the rule to trigger for the given object under consideration. In addition to the 'condition' component, the string component can also be defined in order to give further context to the conditions to be met, such as what to search for in a given object file for condition matching. The string component can be defined as either hexadecimal, text, or regular expressions. Finally, the meta component defines the metadata associated with a given rule. We choose YARA framework to materialize our REGARD's Rules of Engagements (RoEs) as it is popular in the cyber security domain. However, our work introduces an abstract evolutionary syntax notation to capture RoEs. The abstract can then be mapped to any rules-engine-dictated rule syntax. Thus, an organization can choose an existing rules standard if there is a specific rules engine that they are currently using.


\section{REGARD System}
\label{regard-system}


\begin{table}
    \begin{tabularx}{0.47\textwidth} { 
      | >{\centering\arraybackslash}p{0.1\textwidth}|X
      | >{\raggedright\arraybackslash}X | }
        \hline
            \textbf{Symbol} &  \textbf{Meaning} \\
        \hline
            $IRS_{ia}$ & The intermediate action from the IRS AICA-Plan component. \\
        \hline
            $IRS_{t}$ & The target resource (e.g. "/products" webpath or a file system path) of the $IRS_{ia}$ \\ 
        \hline
            $IRS_{s}$ & The source of the $IRS_{ia}$ \\ 
        \hline
            $IRS_{ip}$ & The intermediate plan from the IRS AICA-Plan component containing $\{IRS_{ia}\}$. \\
        \hline
            $IRS_{regard_{input}}$ & The input dictionary, containing the intermediate action, its source, and its managed system target i.e. \{$IRS_{ia}$, $IRS_{s}$, $IRS_{t}$\}\\
        \hline
            $IRS_{regard_{output}}$ & The evaluated action of $IRS_{regard_{input}}$ i.e.  \{$IRS_{a}$\} \\
        \hline
            $IRS_{a}$ & The RoE evaluated final action corresponding of  $IRS_{ia}$ \\
        \hline
            $IRS_{p}$ & The final plan from the IRS REGARD component containing $\{IRS_{a}\}$. \\
        \hline
            $RoE_{r}$ & A generic RoE rule. \\
        \hline
            $RoE_{r}^{i}(j)$ & A specific RoE rule, $j$, of a specific managed system, $i$. \\
        \hline
            $RoE$ & Set of all rules i.e. $\{RoE_{r}\}$ \\  
        \hline
            $r_{a}$ & Antecedent of a $RoE_{r}$. \\ 
        \hline
            $r_{p}$ & Precedent of a $RoE_{r}$. \\ 
        \hline
            $r_{id}$ & A unique id for the rule $RoE_{r}$. \\ 
        \hline
            $r_{s}$ & The anticipated source type $RoE_{r}$ (e.g. username or IP) \\ 
        \hline
            $r_{v}$ & The $IRS_{ia}$ intended action (e.g. create, read, update, delete) \\
        \hline
            $r_{scope}$ & The target scope of $IRS_{ia}$ \\
        \hline
            $r_{c}$ & The constraint applied to the $IRS_{ia}$ (e.g. allow, deny) \\  
        \hline
            $r_{a}$ & The RoE evaluated action corresponding to $IRS_{ia}$ \\
        \hline
            $r_{f}$ & An optional handler to handle complex and custom RoE evaluation for $IRS_{ia}$ \\
        \hline
            $IRS_{D}^i$ & It is a standard deny action for a managed system for example, for a 'denied' constraint rule. \\
        \hline
            $RoE_{t}$ & A meta template that each managed system template \{$RoE_{t}^{i}$\} follows and ultimately all $RoE_{r}$ are compliant to.\\
        \hline
            $RoE_{t}^{i}$ & A template for a specific managed system that each $RoE_{r}^{1}(j)$ of \{$RoE_{r}$\} follows \\    
        \hline  
    \end{tabularx}
    \caption{REGARD system notations.}
    \label{tbl:notation-table}
\end{table}

\begin{figure*}[ht]
    \centering
    \includegraphics[scale=0.55]{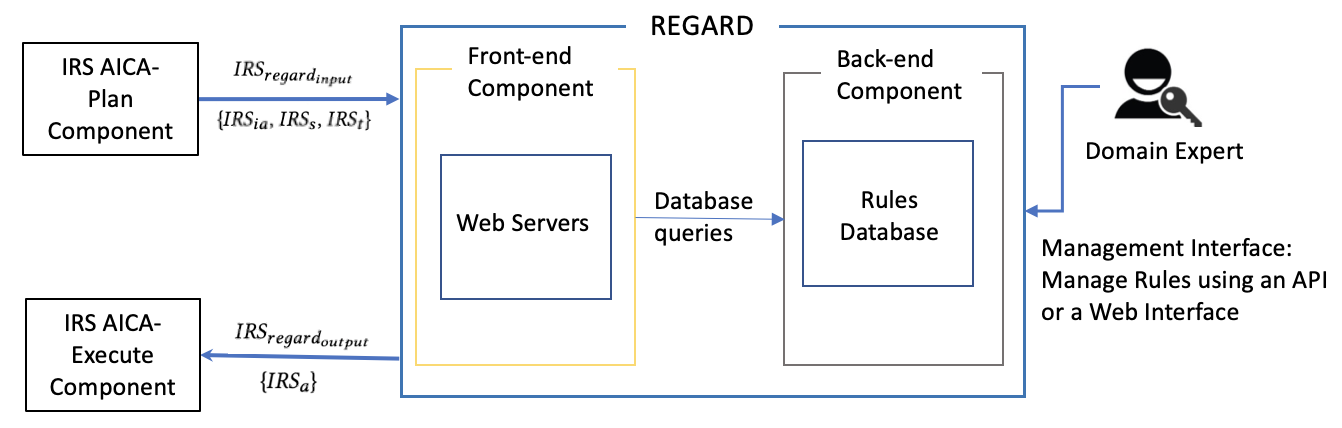}
    \caption{REGARD system components: A system component diagram that contains two components, namely, front-end and back-end. The front-end web component receives the recommended intermediate action from the IRS AICA-Plan components, evaluates the rules stored in the back-end database component, and either allows or denies the intermediate action.}
    \label{fig:regard-system-architecture}
\end{figure*}

This section describes the Rules of EngaGement for Automated cybeR Defense (REGARD) system architecture, design, roles and responsibilities, and rule definitions. Table ~~\ref{tbl:notation-table} summarizes the notations used in this paper. The REGARD component architecture is available in Figure \ref{fig:regard-system-architecture}. The REGARD system extends the Self-Adaptive Autonomic Computing System (SA-ACS) (See Section \ref{self-adaptive-acs}) paradigm and is based on the Automated Intelligent Cyberdefense Agents (AICA) architecture (See Section \ref{aica-architure}).

\noindent\textul{\textit{SA-ACS system information flow based on the AICA architecture with REGARD integration}} (See Figure \ref{fig:regard-system-architecture}): First, the \textit{AICA-Monitoring} managing system components continuously monitor the managed system, thereby collecting the data from AICA sensors and storing the gathered data in the \textit{AICA-Knowledge} components. Second, the \textit{AICA-Analyze} components continuously investigate the gathered data to detect threats. Third, the \textit{AICA-Plan} component creates a plan for each threat, with an aim of attack mitigation. Fourth, the \textit{REGARD component} receives the action set from the plan and creates a final action based on configured Rules of Engagements (RoEs). Fifth, the REGARD component sends the final action to the \textit{AICA-Execute} component. Finally, the AICA-Execute component executes the final action on the managed system through AICA actuators. 

The AICA-Monitor and the AICA-Analyze components constitute an Intrusion Detection System (IDS) that detects a cyber security threat. The AICA-Plan, REGARD, and AICA-Execute component together function as an Intrusion Response System (IRS). These three components mitigate the threat, evaluate each action in the plan against a rule set, formulate new actions, and execute the actions on the managed system.  

REGARD's objective is to enhance managed system security by preventing harmful IRS actions. Its primary role is to operate as an additional security barrier for the intermediate actions computed by the AICA-Plan components. REGARD plays this role in two steps: first, by evaluating them against a set of rules configured by organizational administrators and then by modifying the intermediate actions with a new set of actions. Rules of Engagement Section \ref{dnd}, describes the semantics of both these steps. 

Next, we describe in detail the system design, user roles, rule definitions, execution, management, and operations. This will be followed by details about the REGARD system implementation in the Section \ref{regard-implementation}.

\subsection{System Design and User Roles}
\label{regard-system-design}
Here we detail the REGARD system components and the user roles. 

\subsubsection{System Components}
Different REGARD components and their interactions have been showcased in Fig.~~\ref{fig:regard-system-architecture}. REGARD has two major components: a front-end and a back-end, following a layered architecture pattern to support flexibility and scalability \cite{richards2015software}. 

\noindent{\textul{\textit{The Front-end Component:}}} Its objective is to evaluate the rules and formulate a set of final actions. It is the only component that interacts with the other two SA-ACS IRS components: AICA-Plan and AICA-Execute (Fig.~\ref{fig:regard-architecture}). Moreover, it provides conflict resolution functionality when the input matches the antecedent of more than one rule. We explain the rules and the conflict resolution process in Section ~\ref{dnd}. In addition, the front-end component uses Regular Expressions (RegEx) for rule pattern matching. We describe this further in Section ~\ref{rules-execution}. The front-end component consists of web servers that accept a RESTful API call, process the input, evaluate the rules database situated in the back-end component, retrieve a list of potential final actions, and send the list to the AICA-Execute component.  

\noindent{\textul{\textit{The Back-end Component:}}} This is a rule database. It enables rule operations such as create, read, update, and delete through a management interface provided by the front-end component. Thus, the back-end component does not directly {expose any interface to interact with the database, nor does it} interact with the other AICA components.    

The separate components, following the layered architecture pattern, have the following advantages being able to: 1) scale independently, 2) secure the rules by not exposing the back-end component to the other AICA system 3) provide flexibility to manage them independently. 
We explain how the REGARD system utilizes these advantages in Section ~\ref{REGARD-rule-management-and-operation}.

As shown in Figure \ref{fig:regard-system-architecture}: The AICA-Plan component initiates the REGARD system process by sending a RESTful API request for each of the proposed intermediate actions,  $IRS_{ia}$, present in the intermediate plan, $IRS_{ip}$. The input, $IRS_{regard_{input}}$ consists of the intermediate action ($IRS_{ia}$), its source ($IRS_s$), and the intended target managed system ($IRS_t$). REGARD evaluates the input with a rule set stored in the back-end component. It then formulates the final plan, $IRS_p$, containing the actions, $\{IRS_a\}$. It then creates an output, $IRS_{regard_{output}}$ containing the plan, $IRS_p$, and sends it to the AICA-Execute component. In addition, the AICA-Plan and the REGARD component follow an asynchronous `fire-and-forget' API call pattern while sending the data \cite{voelter2003patterns}. The pattern enables each component to increase the throughput of the API requests by reducing the round trip time. 

In addition, REGARD also provides \textit{two interfaces to manage} REGARD rules. These include a RESTful API interface and a web interface. The former enables system-to-system integration, which helps automate rule management. This can be utilized primarily for batch upload and version control. The latter provides a visual tool to manage the rules interactively. The web management interface can also be utilized to seek administrator input. This is vital in those scenarios where a `human-in-the-loop' intervention is required. 

\subsubsection{System User Roles}\label{sur}
We define two user roles in REGARD to achieve functionality goals and organizational integration. These are 1) \textit{system administrator} and 2) \textit{domain expert}. 

\noindent\textit{\textul{System Administrator User Role:}} The {system administrator} is responsible for the REGARD system operation and management, such as provisioning the deployment infrastructure and deploying the system. 

\noindent\textit{\textul{Domain Expert User Role:}} The {domain expert} is accountable for creating and validating the REGARD rules. They are considered to be experts in the business domain, local organizational cybersecurity posture, functionalities, and the security needs of the managed system. Experts utilize the local organizational knowledge to create/update rules through the web management system interface.

\subsection{Rules of Engagement (RoE) Definition and Design}\label{dnd}

REGARD holds a set of Rules of Engagement (RoE) to protect the managed system according to the instructions provided by the domain expert (See Section \ref{sur}). These rules help limit the action of the IRS on the managed system in compliance with the recommendations of the domain expert. For example, based on domain expert-defined rules,  the REGARD system can prevent an AICA planned action from deleting or modifying a Windows OS registry key, changing browser preferences or bookmarks, modifying an OS DNS or /etc/hosts file, dropping or altering a database table by a rogue user or system, etc. The flexible RoE rule definition and design process in REGARD can represent several security rules to prevent certain actions formulated by the AICA-Plan component from execution on the managed system. The creation of RoE rules depends on the local organizational security policy, local network and compute infrastructure setup, specifications of the managed system, business needs and requirements, etc. For example, the AICA-Plan component produces an intermediate action to delete a file, but the organizational security policy could deny that action. 

\noindent For RoEs, we have taken a few \textit{design decisions} which include: 
\begin{itemize}
    \item We follow a {rule-based expert system definition for REGARD RoE rules} \cite{liu2017fuzzy}. A RoE rule is represented as a set of Left Hand Side (LHS) conditions, called `\textit{antecedents}'. When these antecedents are satisfied, REGARD computes the Right Hand Side (RHS), called `\textit{predicate}'. This is shown in Equation ~\ref{eq1} which is also expressed as an IF...THEN... statement, 
where $RoE_r$ is a single RoE rule with $r_{antecedent}$ and $r_{precedent}$. `\textit{RoE}' is the set of all $RoE_r$.

\begin{fleqn}[\parindent]
\begin{equation} \label{eq1}
    \begin{array}{ll}
         LHS => RHS \\
         RoE_{r}: r_{antecedent} => r_{precedent} \\
    or,\ RoE_{r}: IF\ r_{antecedent}\ THEN\ r_{precedent} \\
    RoE = \{RoE_{r}\}        
    \end{array}
\end{equation} 
\end{fleqn}

\noindent\textit{\textul{Constraint Rule:}} We define a `\textit{constraint rule}', $RoE_r$, that can alter an intermediate action, $IRS_{ia}$, generated by the AICA-Plan component. The antecedent, $r_a$, contains the overwrite instruction configured by the domain expert. 
The rules define three \textit{action types}: 1) those that can be fully automated, 2) those that require a human operator's confirmation, and 3) those that must never be performed. We define the final action, $IRS_a$, as one of three types: the same intermediate action, a modified action, or denied action. 

    \item We employ a \textit{rule allow list} approach. Thus REGARD doesn't permit any intermediate action, $IRS_{ia}$, that doesn't adhere to any rule in RoE. Thus, by default, REGARD does not allow any action on the managed system unless a domain expert configures a matching RoE to handle that action. This design decision is more secure compared to the rule block list alternative. However, this design decision does put an additional burden on the domain expert while creating and maintaining rules for all possible actions. 

    \item We adopt a \textit{template-based approach} for defining the RoE constraint rules to streamline the rule creation process. Each rule adheres to a specific '\textit{managed system template}' that provides a syntax and defines valid types. For example, the action rule part of a web-based managed system should be an HTTP verb. Furthermore, we define a `\textit{meta-template}' that states the syntax for each managed system template and, therefore, each rule. For example, an action rule part of a managed system template is an action type from a set of create, read, update, and delete operations. Or in other words, the meta-template defines the rule syntax and different rule parts but doesn't define any concrete valid values for some rule parts. As another example, a meta-template defines the action rule part to be a verb of type: create, read, update, and delete, while a specific web-managed system template defines the action rule part as an HTTP verb of types: POST, GET, PUT, DELETE, where each of them is a concrete type corresponding to the meta-template action type. Thus, a domain expert follows the meta-template syntax to create a managed system template.  

\end{itemize}

Next, we explain our approach to design the meta-template, and the managed-system template, using symbolic notation. The REGARD system implementation and example case studies are available in Section \ref{regard-implementation}. 

\subsubsection{Meta-template design}
\label{meta-template}
We represent a `\textit{meta template}', $RoE_{t}$, as a set of tuples with $rt_{antecdenent}$ and $rt_{precedent}$, as shown in the Equation ~\ref{eq4}, where antecedent ($LHS$) and precedent ($RHS$) of a rule, ${RoE_r}$, as explained in the Equation \ref{eq1}.

\begin{fleqn}[\parindent]
\begin{equation} \label{eq4}
    \begin{array}{ll}
        RoE_{t} = \{rt_{antecedent}, rt_{precedent}\}\ \textrm{where} \\
        \{rt_{antecedent} => rt_{precedent}\}
    \end{array}
\end{equation} 
\end {fleqn}

The domain experts follow the meta-template to create managed system templates,  $RoE_{t}^i$s. Each managed system template, $RoE^i$, defines a rule set, $RoE_{r}^i (j)$. Thus, there are $\{RoE^i\}$, $\forall i = n$, where $n$ is the number of managed systems. In addition, for each ${RoE^i}$ template, there are $\{RoE_{r}^i (j) \}$, $\forall j=m$, where $m$ is the number of rules for managed system $i$. We define managed system template in Section \ref{system-template}. In addition, we demonstrate in Section \ref{case-studies}, two managed systems, where $n = 2$, with managed system templates $\{RoE^1, RoE^2\}$ for each system, and each system template has two rules, i.e. $m = 2$, thus $RoE=\{RoE_{r}^1 (1), RoE_{r}^1 (2), RoE_{r}^2 (1), RoE_{r}^2 (2)\}$.

\begin{fleqn}[\parindent]
\begin{equation} \label{eq5}
    \begin{array}{ll}
        RoE_{t} = \{rt_{id}, rt_{s}, rt_{v}, rt_{scope}, rt_{c}, rt_{a}, rt_{h}\}\ \textrm{where} \\
        rt_{id}\ :\ A\ `Unique\ ID' \textrm{ of a particular rule $RoE_{r}$}, \\
        rt_{s}\ :\ The\ `source\ type' \\
        \qquad \textrm{ of the action $IRS_{ia}$ of the rule $RoE_{r}$}, \\ 
        rt_{v}\ :\ The\ `intermediate\ action\ type' \\
        \qquad \textrm{ of the action $IRS_{ia}$ of the rule $RoE_{r}$}, \\ 
        rt_{scope}\ :\ The\ `resource\ scope' \\
        \qquad \textrm{ of the action $IRS_{ia}$ of the rule $RoE_{r}$}, \\ 
        rt_{c}\ :\ The\ `constraint\ type' \\
        \qquad \textrm{ of the action $IRS_{ia}$ of the rule $RoE_{r}$}, \\ 
        rt_{a}\ :\ The\ `final\ action' \\
        \qquad \textrm{ of the action $IRS_{a}$ of the rule $RoE_{r}$}, \\ 
        rt_{h}\ :\ An\ optional\ `Handler' \\
        \qquad \textrm{ of the rule $RoE_{r}$} 
    \end{array}
\end{equation} 
\end{fleqn}

Thus, we represent $rt_{antecedent}$ and $rt_{precedent}$ as two subsets of the $RoE_{t}$ tuple, as shown in equation ~\ref{eq6}.

\begin{fleqn}[\parindent]
\begin{equation} \label{eq6}
    \begin{array}{ll}
         rt_{antecedent} = \{t_{id}, rt_{s}, rt_{v}, rt_{scope}\}\ \textrm{and} \\
         rt_{precenent} = \{rt_{c}, rt_{a}, rt_{h}\}   \\
         \textrm{Thus, } \\
         \{t_{id}, rt_{s}, rt_{v}, rt_{scope}\}\ => \{rt_{c}, rt_{a}, rt_{h}\}
    \end{array}
\end{equation} 
\end{fleqn}

\noindent Next, we define each term of the $RoE_{t}$ tuple from the Equation ~\ref{eq6}.

\noindent $rt_{id}$ is a unique id creator for a role $RoE_{r}$ following a template as shown in the Equation ~\ref{eq7}. It contains a concatenated abbreviated category, and one or more subcategories separated by a hyphen. 

\begin{fleqn}[\parindent]
\begin{equation} \label{eq7}
     rt_{id} \textrm{ } = \textrm{ }<Category-subcategory[-subcategory]>
\end{equation} 
\end {fleqn}

The definition of an $rt_{id}$ along with a description is stored in a dictionary in the format shown in equation ~\ref{eq8}. 

\begin{fleqn}[\parindent]
\begin{equation} \label{eq8}
    \begin{array}{ll}
    \ Action\ ID: \\
    <Category-subcategory[-subcategory], description>\\
\textrm{e.g.} \\
\textrm{\{"id":"NET-L4-DDOS", "description":"DDos Attack."\}}\\
\textrm{\{"id":"WEB-FE-SQL", "description":"Layer 7 SQL Injection."\}}\\
where \\
\textrm{Category: Network, WebApp} \\
\textrm{Subcategories:} \\
\textrm{Network: -> Layer 7 -> DDoS} \\
\textrm{WebApp -> DB -> SQL Injection}
    \end{array}
\end{equation}
\end{fleqn}

\noindent $rt_{s}$ is the source type of an intermediate action $IRS_{ia}$ and defines the source categories. As an illustration, $rt_{s}$ with two potential source types, namely, `\textit{username}' or `\textit{IP}' as shown in equation ~\ref{eq9}.

\begin{fleqn}[\parindent]
\begin{equation} \label{eq9}
     rt_{s} = (username|IP)
\end{equation} 
\end {fleqn}

\noindent $rt_{v}$ is the action type on the managed system. For example it could four types: `\textit{create}', `\textit{read}', `\textit{update}, or `\textit{delete}' as shown in equation ~\ref{eq10}.

\begin{fleqn}[\parindent]
\begin{equation} \label{eq10}
     rt_{v} = (create|read|update|delete)
\end{equation} 
\end {fleqn}

\noindent $rt_{scope}$ is the scope of target of the action $IRS_{ia}$. As an example, the scope is similar to $rt_{t}$.

\noindent $rt_{c}$ is the constraint applied to the intermediate action $IRS_{ia}$ is it matches $RoE_{r}$. For example, it could be one of the following: `\textit{allow}', `\textit{deny}' or `\textit{allowWithLog}' as shown in equation ~\ref{eq11}. Allow permits $IRS_{rt}$ to be executed on the managed system and is considered as `\textit{passthrough}'. Deny does not allow the $IRS_{ia}$. Finally, allowWithLog permits the action to pass through but only after logging the action for an admin for further analysis.

\begin{fleqn}[\parindent]
\begin{equation} \label{eq11}
    \begin{array} {ll}
    rt_{c} = (allow|deny|allowWithLog) \\
    where \\
    IF\ (allow)\ THEN\ (IRS_p\{IRS_ia\}) \\
    \qquad \textrm{i.e.}\ IRS_a = IRS_ia \\
    IF\ (deny)\ THEN\ (IRS_p\{IRS_D\}) \\
    IF\ (allowWithLog)\ THEN\ (IRS_p\{IRS_ia, log\ message\}) \\
    \end{array}    
\end{equation} 
\end {fleqn}

\noindent $rt_{ta}$ defines the output action after evaluating with RoE. For example, it could have three corresponding actions based on the $rt_{c}$ type, namely: ${IRS_{ia}}$, ${IRS_D}$, ${IRS_{ia}, log message}$ where $IRS_{D}$ is a standard deny action relevant to the respective managed system as shown in equation ~\ref{eq12}.

\begin{fleqn}[\parindent]
\begin{equation} \label{eq12}
    \begin{array} {ll}
    IRS_D:\textrm{a standard deny action for the managed system} \\
    \textrm{e.g.} \\
    \textrm{\textbf{404} for a managed web\ system}\\
    \qquad \textrm{of a}\ DELETE\ \textrm{/home/index.html $IRS_{ia}$ action} \\
    \textrm{or} \\
    \textrm{\textbf{CLOSED} for a managed network infrastructure management} \\ 
    \qquad \textrm{system of a}\ \textrm{SYN $IRS_{ia}$ action}\ \\
    \qquad \textrm{to a DNS server from an external IP.}\ \\
    \end{array}    
\end{equation} 
\end {fleqn}

\noindent $r_{h}$ is a `\textit{handler}' which provides additional rules for further complex processing of the input $IRS_{ia}$, evaluating $RoE_{r}$, or creating the final $IRS_{a}$.

Next, we define managed system templates, one for each managed system, that adhere to the meta template, $RoE_t$. 

\subsubsection{Managed system-template design} 
\label{system-template}
We define a `\textit{managed system template}' for each managed system, $RoE^{i}$, that adheres to the meta template definition of $RoE_{t}$ (shown in equation ~\ref{eq5}) in Section ~\ref{meta-template}. A few examples of $RoE^{i}$ include a web-managed system ($RoE^1$), a file-managed system ($RoE^2$), and a network infrastructure management-managed system ($RoE^3$.) Each $RoE^{i}$ adheres to the meta-template definition. For example, a $RoE^i$ defines an action rule part complying with the meta-template action rule part. For example, a web-managed system template, $RoE^1$, defines the action in the rule to have an HTTP verb value (POST, GET, PUT, DELETE) based on the meta-template action definition action type, $rt_{v}$, (create, read, update, or delete operation). We demonstrate two managed system templates in Section ~\ref{case-studies}.

\subsection{Rules Execution and Conflict Resolution}
\label{rules-execution}
Once the RoE rules have been defined (See Section ~\ref{dnd}), these are then evaluated by REGARD against the input received from the AICA-Plan component (See figure \ref{fig:regard-system-architecture}). REGARD executes the rules against the AICA-Plan intermediate actions in four steps. 

\begin{enumerate}
    \item The REGARD component receives \{$IRS_{ia}$, $IRS_s$, $IRS_t$\} from the AICA-Plan component as shown in equation ~\ref{re1}. 
    \item The REGARD system applies \{$RoE_{r}$\} to the input, we represent the evaluation engine as $f(IRS_{regard_{input}},$ $RoE_{r})$.
    \item REGARD produces the final actions set, \{$IRS_{a}$\} as shown in equation ~\ref{re2}.
    \item REGARD passes the final action set to the IRS AICA-Execute component following equation ~\ref{eq1}. 
\end{enumerate}

\begin{fleqn}[\parindent]
\begin{subequations}\label{re}
    \begin{equation} \label{re1}
        \begin{array} {ll}
        IRS_{regard_{input}} = \{IRS_{ia}, IRS_s, IRS_t \} \\
        \end{array}    
    \end{equation} 
    \begin{equation} \label{re2}
        \begin{array} {ll}
        f(IRS_{regard_{input}}, RoE_{r}) = IRS_{regard_{output}} \\
        where \\
        IRS_{regard_{output}} = \{ IRS_a\}
        \end{array}    
    \end{equation}     
\end{subequations}
\end {fleqn}

REGARD uses the regular expression to match the input with the LHS RoE part and outputs the RHS rule determined by RoE constraint \{$r_{c}$\}, of $RoE_{r}$ as shown in equation ~\ref{eq5}. A regular expression is a common tool because of its flexibility, and expressive ability \cite{xu2016survey}. Thus, a domain expert can take advantage of the regular expression pattern-matching benefit, a powerful technique that allows domain experts to define the rules concisely and precisely.


While RoEs are being executed, it is possible that more than one RoE matches the intermediate action, $IRS_{ia}$. The resulting `\textit{rule conflict}' needs to be resolved for the REGARD system to function properly. In the REGARD system, our conflict resolution strategy is: \textit{the rule with more restrictive preference takes precedence, and the final action,} $IRS_a$, \textit{corresponding to the restrictive rule is selected.} A managed system determines the order of precedence.

For example, when there are two rules for a managed file system with a template, $RoE^1$, where one rule, $RoE_{r}^1(1)$, is configured with target $r_{t}(1)$ as ``/users``,  action $r_{v}(1)$ as ''delete'', constraint $r_{c}(1)$ as ''deny'', action as $r_{a}(1)$ as ''Permission denied'' and another one rule, $RoE_{r}^1(2)$ is configured with target $r_{t}(2)$ as ''/users/admin'',  action $r_{v}(2)$ as ''delete'', constraint $r_{c}(2)$ as ''allow'', action as $r_{a}(1)$ ''{empty}'' then $IRS_{ia}$ to delete users/admin is not allowed as shown in equation ~\ref{eq13}. 

\begin{fleqn}[\parindent]
\begin{subequations}\label{eq13}
    \begin{equation} \label{eq13a}
        \begin{array} {ll}
        RoE_r^1(1) = \{..., r_t(="/users"), \\
        \qquad r_ia(="delete"), r_c(="deny"), \\
        r_a(="Permission Denied")
        ... \}
        \end{array}    
    \end{equation} 
    \begin{equation} \label{eq13b}
        \begin{array} {ll}
        RoE_r^1(2) = \{..., r_t(="/users/admin"), \\
        \qquad r_ia(="delete"),  r_c(="allow"), \\
        r_a(="")
        ... \}
        \end{array}    
    \end{equation}     
    \begin{equation} \label{eq13c}
        \begin{array} {ll}
        \textrm{RoE evaluation:} \\
        IF\ RoE_r^1(1)\ and\ RoE_r^1(2)\ THEN\ \{r_{a}\}(1)
        \end{array}    
    \end{equation}     
\end{subequations}
\end {fleqn}

\subsection{Rule Management and Operation}
\label{REGARD-rule-management-and-operation}
The REGARD management interface allows a domain expert to create, read, update, or delete the rules. They can do so in two ways to manage the RoE: a web interface or a REST API endpoint. The former provides a visual and interactive mechanism to manage the rules, while the latter offers a programmatic mechanism. 
The API enables bulk management and version control of RoEs. The version management tracks the rules changes over time which aids in the audit process. Next, we primarily discuss the RoE management process and a few REGARD maintenance operations. 

To track the rule changes and to maintain the REGARD system, including recovery in the event of a natural disaster or system crash, the system administrator frequently exports the RoE database and saves them in a secure location. The system administrator saves the database exports following the Recovery Time Objective (RTO) and Recovery Point Objective (RPO) as agreed in Service Level Agreements (SLA). Moreover, the system administrator creates a new REGARD system using the database exports for business continuity when the primary REGARD system is unavailable and unrecoverable. In addition, the system administrator allows access to the management interface and the REGARD administrative interface only from the Intranet for additional security. Or in other words, the management interfaces cannot be accessed outside the REGARD system, such as the AICA-Plan or AICA-Execute components. Finally, the system administrator follows a change management process to update the REGARD system software and infrastructure. Describing the REGARD continuous integration and deployment, or backup and disaster recovery strategy, or other operations used in a software development life-cycle is out of the scope of this paper.

\section{REGARD Implementation}
\label{regard-implementation}

\begin{figure*}[ht]
    \centering
    \includegraphics[width=1\textwidth]{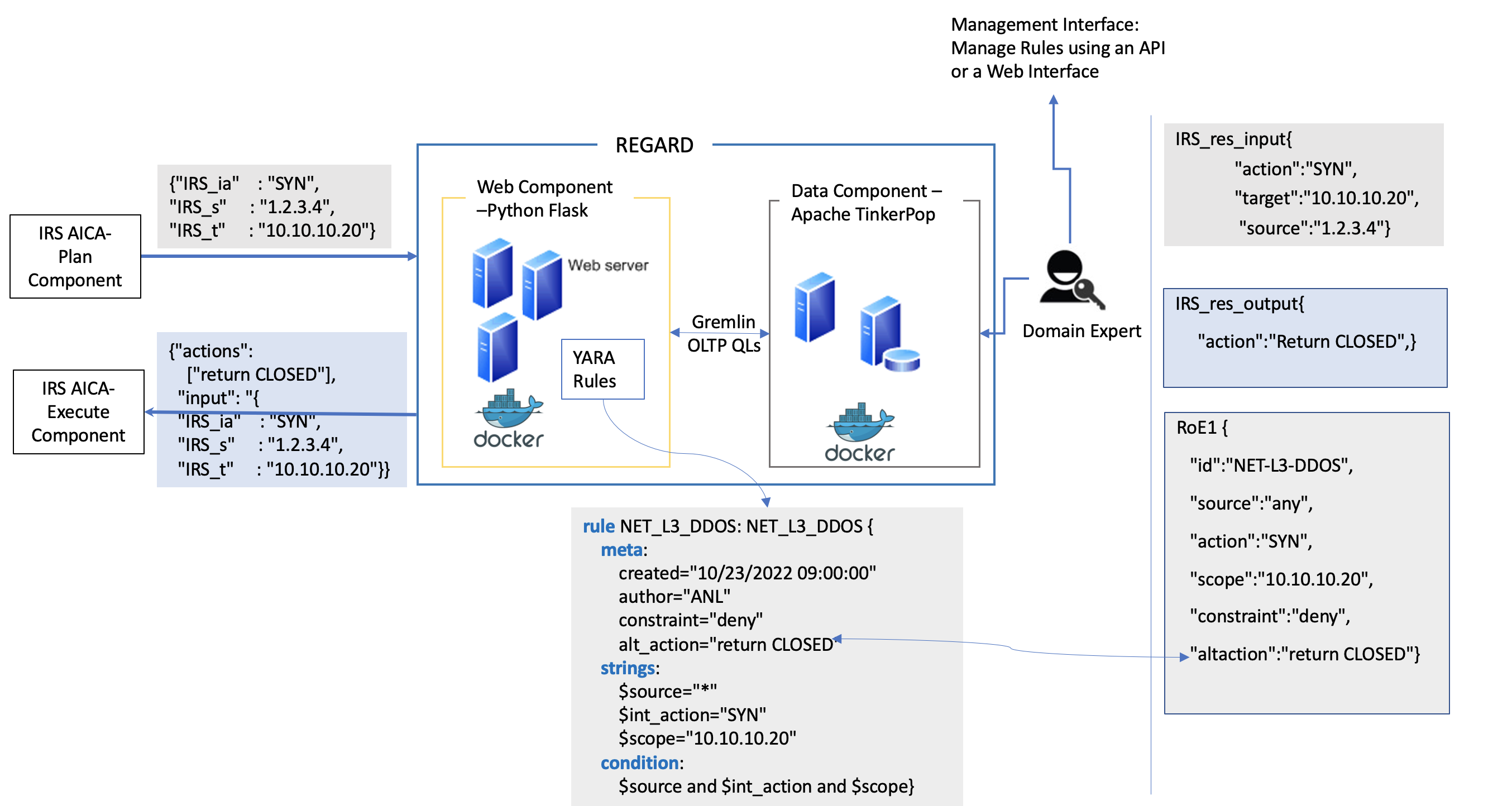}
    \caption{REGARD system implementation: An implementation of the REGARD components and the rules (Rules of Engagement). The front-end web component and the back-end database are deployed as Python Flask and Apache TinkerPop Docker containers, respectively. In addition, the RoEs and the input/output abstract syntax notations are materialized as YARA rules and JSON objects, respectively.}
    \label{fig:regard-system-implementation}
\end{figure*}

This section takes the abstract REGARD design and RoE constrain rules definition and demonstrates their implementation. The system component design is described in Section \ref{regard-system-design}, and the RoE rules are defined in Section \ref{dnd}. Moreover, the Section demonstrates the RoE rules implementation for two managed systems. We choose to implement the system and the rules using certain technologies. However, the system's plug-and-play \cite{abdulrazak2022iot} architecture and the rules permit selecting a different technology set. Thus, the approach showcases the power of the REGARD system design and the core RoE rules semantic definitions.

REGARD follows a micro-service architecture to implement the two components, as shown in Fig. ~\ref{fig:regard-system-architecture}. The Python Flask framework \cite{grinberg2018flask} provides the front-end component functionality. Apache TinkerPop \cite{tinkerpop}, a graph database, provides the back-end component functionality. Both the Python Flask and the TinkerPop software are provided as Docker containers \cite{merkel2014docker}.

REGARD's execution flow is as follows: First, the AICA-Plan component sends an HTTP Request JSON payload with the intermediate action, source, and target ($IRS_{ia}$, $IRS_s$, $IRS_t$) as a dictionary. Second, the REGARD front-end Flask web framework component receives the input ($IRS_{regard_{input}}$). Third, a Python web program parses the input to match with the LHS of the rule set (\{$RoE$\}). Fourth, the Python program queries the REGARD back-end TinkerPop component using Gremlin QLs to evaluate the input with the LHS. Fifth, based on the matching rules, the program formulates an output with the final action set (\{$IRS_a$\}). Finally, the program sends an HTTP request to the AICA-Execute component with a JSON payload of the final actions and the input it received. The REGARD implementation is shown in Fig. ~\ref{fig:regard-system-implementation}. REGARD components can be independently scaled and managed as explained in Section ~\ref{regard-system-design}. 

In addition, REGARD also uses YARA rules \cite{yara} to implement the rules' definition and execution as explained in sections ~\ref{dnd} and ~\ref{rules-execution}. YARA is a popular semantic language among malware researchers \cite{naik2020evaluating, naik2019cyberthreat, naik2019augmented}. It provides a syntactic framework to describe a malware family based on textual or binary patterns. Our use of YARA further aids in the integration and adoption of the REGARD system in organizations that use this popular framework for cyber operations. 
We implement the rule, $RoE_{r}$, that complies with $RoE_{t}$, represented in equation ~\ref{eq4}, following the YARA specifications.

\begin{fleqn}[\parindent]
\begin{equation} \label{eq14}
    RoE_{r} = \{r_{id}, r_{s}, r_{v}, r_{scope}, r_{c}, r_{a}, r_{h}\}
\end{equation} 
\end {fleqn}

The above $RoE_{r}$, based on the meta template definition as described in Section ~\ref{dnd}, is materialized in three parts using the YARA specifications. YARA is a popular tool that malware researchers use to classify malware semantically in a text format. The first part is the $RoE_{r}$ subset $\{r_{id}, r_{s}, r_{v}, r_{scope} \}$ that is defined as a YARA `\textit{rule}'. The second part is the subset $\{r_{c}, r_{a}\}$ that is represented in the `\textit{meta}' section of the YARA rule. Finally, the third part is the subset $\{r_{h}\}$ that is represented as a `\textit{callback}'. The three parts are shown in equation ~\ref{eq15}. The third part provides a mechanism to evaluate complex rules matching logic that need rule processing beyond the matching capabilities offered by the first part of the rule. Thus, this part is an optional rule part. 

\begin{fleqn}[\parindent]
\begin{subequations}\label{eq15}
    \begin{equation} \label{eq15a}
        \begin{array} {ll}
        \textrm{\{}{r_{id}, r_{s}, r_{v}, r_{scope}} \textrm{\}} 
        \textrm{ as a YARA `\textit{rule}'.}
        \end{array}    
    \end{equation} 
    \begin{equation} \label{eq15b}
        \begin{array} {ll}
        \textrm{\{}{r_{c}, r_{a}, r_{h}} \textrm{\}}  
        \textrm{ as a YARA `\textit{meta rule}' set}.
        \end{array}    
    \end{equation}     
    \begin{equation} \label{eq15c}
        \begin{array} {ll}
        \textrm{\{}{r_{h}} \textrm{\}}  
        \textrm{ as a YARA `\textit{callback} function'}.
        \end{array}    
    \end{equation}     
\end{subequations}
\end {fleqn}

The mapping of the first part of the $RoE_{r}$ as mentioned in equation ~\ref{eq15a} represented as a YARA `\textit{rule}' and is shown in listing~~\ref{list:RoE-part1}.
\begin{lstlisting} [language=YARA, mathescape=true, caption=First part of the \textit{$RoE_{r}$} as shown in the Equation ~\ref{eq15a}., label=list:RoE-part1]
rule $r_{id}$ {
    meta:
        created     = "date"
        author      = "organization name"
    strings:
        $\$$source      = "$r_{s}$"
        $\$$int_action  = "$r_{v}$"
        $\$$scope       = "$r_{scope}$"
    condition: 
        $\$$source and  $\$$int_action and $\$$scope}
\end{lstlisting}

The mapping of the second part of the $RoE_{r}$ as mentioned in equation ~\ref{eq15b}, is represented in the `\textit{meta}' section of the YARA rule and is shown in listing~~\ref{list:RoE-part2}.
\begin{lstlisting} [language=YARA, mathescape=true, caption=Second part of the \textit{$RoE_{r}$} as shown in the Equation ~\ref{eq15b}., label=list:RoE-part2]
rule $r_{id}$ {
    meta:
        ...
        'constraint'= '$r_{c}$', 'alt_action'= '$r_{a}$' 
        ...} 
\end{lstlisting}

The mapping of the third part of the $RoE_{r}$ as mentioned in the Equation ~\ref{eq15c}, is represented as YARA `\textit{callback function}' and is shown in listing~~\ref{list:RoE-part3}.
\begin{lstlisting} [language=YARA, mathescape=true, caption=Third part of the \textit{$RoE_{r}$} as shown in the Equation ~\ref{eq15c}., label=list:RoE-part3]
matches = rules.match(
    '/foo/bar/my_file',
    callback = $r_{h}$)
\end{lstlisting}

\subsection{REGARD Case Studies for Cyber Defense}
\label{case-studies}

In this section, we demonstrate \{$RoE_r$\} and its YARA implementation for two managed systems, namely, \textit{a web system} and \textit{a network infrastructure management system}. First, we use the rule's symbolic notations to exemplify the use of REGARD. 
Next, we define one template for each of the two managed systems. Both the templates comply with the managed system template as explained in Section ~\ref{system-template} and the meta template definition described in Section ~\ref{meta-template}. Finally, we demonstrate two rules for each managed system and showcase their execution for two AICA Plan Component inputs as explained in Section ~\ref{rules-execution}.

\subsubsection{Case Study 1 - Managed Web System}
\label{example1}
In the first case study, we define a template for a managed web system as $RoE^{1}$, as shown in equation ~\ref{eq16}, where the AICA-Plan Component's planned action is one of the HTTP methods \cite{httpverb}. Moreover, REGARD overwrites the constraint actions with a deny action that is defined as a `404' error. 
To demonstrate the $RoE^1$, we define two rules following the semantics of Section \ref{dnd}, define them in YARA syntax, and show their evaluation for two inputs from the AICA-Plan Component, adhering to the rules execution procedure as explained in Section ~\ref{rules-execution}. In the first rule, the REGARD system interjects, per the domain experts' laid out rules, and constrains an AICA-Plan intermediate action from deleting any web content from the system. The second rule constrains the intermediate action from accessing any sensitive information stored in the `/admin/' directory or any of its subdirectories.

\begin{fleqn}[\parindent]
\begin{equation} \label{eq16}
    \begin{array}{ll}
        RoE^1 = \{rt_{id}, rt_{s}, rt_{v}, rt_{scope}, rt_{c}, rt_{a}, rt_{h}\}\ \textrm{where} \\
        rt_{id},\ rt_{s},\ rt_{scope},\ rt_{c},\ and\ \ rt_{h}\ \\
        \qquad \textrm{are same as described in the Equation}\ ~\ref{eq5},\\
        rt_{v}:\ \textrm{is an HTTP Verb \cite{httpverb}, e.g. GET/POST/PUT/DELETE}, \\ 
        IRS_{D}^1\ :\ \textrm{Resource Not Found - 404 Error Code.} \\
        \textrm{where} \\
        IRS_D \textrm{ is a standard deny action for the managed} \\
        \textrm{system as shown in the equation \ref{eq12}.} 
    \end{array}
\end{equation} 
\end {fleqn}

\noindent\textul{Rules for Case Study 1:} We define two rules for a web-managed system to showcase their representation following the symbolic notation that we introduced in Section ~\ref{dnd} and represented in equation ~\ref{eq5}, as shown in equation ~\ref{eq16}.

\begin{itemize}
\item Rule 1:
\label{example1-rule1}
This RoE rule specifies that the IRS AICA-Plan intermediate action to delete any resource from the website originating from any source IP should be constrained. Based on the rule, REGARD interjects and overwrites the delete intermediate action with a deny action. Thus, $RoE_r^1(1)$ is a specific rule that `\textit{denies}' ($r_{c}$(=''deny'')), any requests for`\textit{any source}' ($r_{s}$ (=any)), to `\textit{delete}' ($r_{v}$(=''DELETE'')) a scope `\textit{/}' ($r_{scope}$(=''/'')) and returns an error code `\textit{404}' ($IRS_{a}$ = $IRS_{D}^1$) as shown in listing ~\ref{list:web-system-rule1}, following the rules (Section ~\ref{meta-template}).

\begin{lstlisting} [language=YARA, mathescape=true, caption=$RoE_{r}^1(1)$ definition., label=list:web-system-rule1] 
{ "$r_{id}$"       : "WEB-FE-XSS-1",
  "$r_{s}$"        : "any", 
  "$r_{v}$"        : "DELETE",
  "$r_{scope}$"    : "/",
  "$r_{c}$"        : "deny",
  "$r_{a}$"        : "return 404"}
\end{lstlisting}

The YARA specification of the $RoE_r^1(1)$ shown in listing \ref{list:web-system-rule1} consists of two parts as per the specification in equation \ref{eq15} and are shown in listing ~\ref{list:web-system-rule1-yara1} and ~\ref{list:web-system-rule1-yara2} respectively.

\begin{lstlisting} [language=YARA, mathescape=true, caption=YARA Spec for the first part of the \textit{$RoE_r^1(1)$}., label=list:web-system-rule1-yara1]
rule WEB-FE-XSS-1 {
    meta:
        created     = "10/23/2022 09:00:00"
        author      = "ANL"
        
    strings:
        $\$$source      = "*"
        $\$$int_action  = "DELETE" 
        $\$$scope       = "/"
        
    condition: 
        $\$$source 
            and $\$$int_action and $\$$scope }
\end{lstlisting}

\begin{lstlisting} [language=YARA, mathescape=true, caption=YARA Spec for the second part of the \textit{$RoE_r^1(1)$}., label=list:web-system-rule1-yara2]
rule WEB-FE-XSS-1 {
    meta:
        ...
        '$constraint$'= 'deny', '$alt\_action$'= 'return 404'... } 
        

\end{lstlisting}

\item Rule 2:
\label{example1-rule2}
This RoE rule constrains the IRS AICA-Plan intermediate action to access any content under the `/admin' directory. REGARD system overwrites the intermediate action with a final deny action. Thus, $RoE_r^1(2)$ is a specific rule that `\textit{denies}' ($r_{c}$(="deny")), any requests for `\textit{any source} ($r_{s}$ (=*)), to `\textit{read}' ($r_{v}$(="GET")) a scope `\textit{/admin}' ($r_{scope}$) and returns an error code `\textit{404}' ($IRS_{a}$ = $IRS_{D}^1$) as shown in listing ~\ref{list:web-system-rule2}.

\begin{lstlisting} [language=YARA, mathescape=true, caption=$RoE_{r}^1(2)$ rule., label=list:web-system-rule2]
{ "$r_{id}$"       : "WEB-FE-XSS-2",
  "$r_{s}$"        : "any", 
  "$r_{v}$"        : "GET",
  "$r_{scope}$"    : "/admin",
  "$r_{c}$"        : "deny",
  "$r_{a}$"        : "return 404"}
\end{lstlisting}

The YARA specification of the $RoE_r^1(2)$ shown in listing in \ref{list:web-system-rule2} consists of two parts as per the specification in equation \ref{eq15} and are shown in listing ~\ref{list:web-system-rule2-yara1} and ~\ref{list:web-system-rule2-yara2} respectively.

\begin{lstlisting} [language=YARA, mathescape=true, caption=YARA Spec for the first part of the \textit{$RoE_r^1(2)$}., label=list:web-system-rule2-yara1]
rule WEB-FE-XSS-2 {
    meta:
        created     = "10/23/2022 09:00:00"
        author      = "ANL"
    strings:
        $\$$source      = "*"
        $\$$int_action  = "GET" 
        $\$$scope       = "/admin"
        
    condition: 
        $\$$source and  
            and $\$$int_action and $\$$scope}
\end{lstlisting}

\begin{lstlisting} [language=YARA, mathescape=true, caption=YARA Spec for the second part of the \textit{$RoE_r^1(2)$}., label=list:web-system-rule2-yara2]
rule WEB-FE-XSS-2 {
    meta: 
    ...
    '$constraint$'= 'deny', '$alt\_action$'= 'return 404'...}
\end{lstlisting} 
\end{itemize}

\noindent\textul{Inputs for Case Study 1:}
We take two AICA-Plan Component inputs containing intermediate action and demonstrate the two RoE rules' evaluation mechanism. We showcase the pattern-matching technique of the rules execution process, as described in Section ~\ref{rules-execution} for a REGARD input. 

\begin{itemize}
\item Input 1:
\label{example1-input1}
The AICA-Plan component plan is set to let a source(`1.2.3.4.')  `delete' the `/products' content. However, Rule 1 constrains the planned intermediate action and denies it. The REGARD execution steps are as follows: REGARD receives input from the AICA-Plan component. The input, $IRS_{regard_{input}}^1(1)$, content is shown in listing ~\ref{list:web-system-irs-input-1}. The input semantics adheres to the definition shown in equation ~\ref{re1}. Then, REGARD applies \{$RoE_{r}^1$\} to the input as mentioned in equation ~\ref{re2}. Next, it iterates through the rule set and finds a matching rule, $RoE_r^1(1)$. Finally, it overwrites the intermediate action with a constraint action, which is an Error, \{$IRS_D^1$\} as $IRS_{regard_{output}}^1(1)$ following the Equation ~\ref{re2} as Listed in ~\ref{list:web-system-irs-output-1}.

\begin{lstlisting} [language=YARA, mathescape=true, caption=$IRS_{regard_{input}}^1(1)$ input., label=list:web-system-irs-input-1]
{ "$IRS_{ia}$"       : "DELETE",
  "$IRS_{s}$"        : "1.2.3.4",   
  "$IRS_{t}$"        : "/products" }
\end{lstlisting}

\begin{lstlisting} [language=YARA, mathescape=true, caption=$IRS_{regard_{output}}^1(1)$ output., label=list:web-system-irs-output-1]
{ "$action$"       : "Return 404" }
\end{lstlisting}

\item Input 2:
\label{example1-input2} 
The REGARD input from the IRS Component plan allows the source (`'1.2.3.4') to access the content of `/admin/user.' However, Rule 2 constrains the intermediate action. The flow is similar to the Input 1 demonstration, but for this input, REGARD finds a matching rule, Rule 2. Listings ~\ref{list:web-system-irs-input-2} and ~\ref{list:web-system-irs-output-2} show the $IRS_{regard_{input}}^1(2)$ and the $IRS_{regard_{output}}^1(2)$ respectively.



\begin{lstlisting} [language=YARA, mathescape=true, caption=$IRS_{regard_{input}}^1(2)$ input., label=list:web-system-irs-input-2]
{ "$IRS_{ia}$"       : "GET",
  "$IRS_{s}$"        : "1.2.3.4", 
  "$IRS_{t}$"        : "/admin/user"}
\end{lstlisting}

\begin{lstlisting} [language=YARA, mathescape=true, caption=$IRS_{regard_{output}}^1(2)$ output., label=list:web-system-irs-output-2]
{ "$action$"       : "Return 404" }
\end{lstlisting}
\end{itemize}

\subsubsection{Case Study 2 - Network Infrastructure Management System}
\label{example2}
In the second case study, we define a template for a managed network management system as $RoE^{2}$ as shown in the Equation ~\ref{eq17}, that uses TCP state transition actions \cite[Section18.6]{fall2011tcp} as a verb and defines `CLOSED' as the final actions for constrains actions as specified by the domain experts. We take a similar approach to Case Study 1 and demonstrate the $RoE^2$ with two rules, define them in YARA syntax, and show their evaluation for two $IRS_{regard_{input}}^{2}$, adhering to the rules execution procedure as explained in Section ~\ref{rules-execution}. In the first rule, the REGARD system constrains an IRS AICA-Plan intermediate action to allow any source to connect to a DNS server. The second rule constrains the intermediate action from amending firewall rules.

\begin{fleqn}[\parindent]
\begin{equation} \label{eq17}
    \begin{array}{ll}
        RoE^2 = \{rt_{id}, rt_{s}, rt_{v}, rt_{scope}, rt_{c}, rt_{a}, rt_{h}\}\ \textrm{where} \\
        rt_{id},\ rt_{s},\ rt_{c},\ and\ \ rt_{h}\ \\
        \qquad \textrm{are same as described in the Equation}\ ~\ref{eq5},\\
        rt_{scope}:\ \textrm{An IP or a CIDR} \\
        rt_{v}:\ \textrm{TCP state transition actions \cite[Section18.6]{fall2011tcp}} \\
        \qquad\textrm{e.g. TCP state transition: SYN/CLOSED}\\
        IRS_{D}^2\ :\ \textrm{Connection CLOSED.} \\
    \end{array}
\end{equation} 
\end {fleqn}

\noindent\textul{Rules for Case Study 2:}
We define two rules for a managed network management system as to showcase their representation following the symbolic notation that we introduced in Section ~\ref{dnd} and represented in the equation ~\ref{eq5}, as shown in the equation ~\ref{eq16}.

\begin{itemize}
\item Rule 1:
\label{example2-rule1}
This rule constrains an IRS AICA-Plan intermediate action to allow any source to connect to a DNS server. Evaluating the rules, REGARD overwrites the intermediate action with a deny action. Thus, $RoE_r^2(1)$ is a specific rule that `\textit{denies}' ($r_{c}$(="deny")), any requests from any `\textit{any source}' ($r_{s}$ (=any)) to `\textit{connect}' ($r_{v}$(="SYN")) create a connection to the DNS with IP `\textit{10.10.10.20}' ($r_{scope}$(="10.10.10.20") and returns a deny action, `\textit{CLOSED}' ($IRS_{a}$ = $IRS_{D}^2$) as shown in listing ~\ref{list:net-management-system-rule1}, following the rules (Section ~\ref{meta-template}).

\begin{lstlisting} [language=YARA, mathescape=true, caption=$RoE_{r}^2(1)$ rule., label=list:net-management-system-rule1] 
{ "$r_{id}$"       : "NET-L3-DDOS",
  "$r_{s}$"        : "any", 
  "$r_{v}$"        : "SYN",
  "$r_{scope}$"    : "10.10.10.20",
  "$r_{c}$"        : "deny",
  "$r_{a}$"        : "return CLOSED"}
\end{lstlisting}

The YARA specification of the $RoE_r^2(1)$ shown in listing \ref{list:net-management-system-rule1} consist of two parts as per the specification in equation \ref{eq15} and are shown in listings ~\ref{list:net-management-system-rule1-yara1} and ~\ref{list:net-management-system-rule1-yara2} respectively.

\begin{lstlisting} [language=YARA, mathescape=true, caption=YARA Spec for the first part of the \textit{$RoE_r^2(1)$}., label=list:net-management-system-rule1-yara1]
rule NET-L3-DDOS {
    meta:
        created     = "10/23/2022 09:00:00"
        author      = "ANL"
        
    strings:
        $\$$source      = "*"
        $\$$int_action  = "SYN" 
        $\$$scope       = "10.10.10.20"
        
    condition: 
        $\$$source 
            and $\$$int_action and $\$$scope }
\end{lstlisting}

\begin{lstlisting} [language=YARA, mathescape=true, caption=YARA Spec for the second part of the \textit{$RoE_r^2(1)$}., label=list:net-management-system-rule1-yara2]
rule NET-L3-DDOS {
    meta:
        ...
        '$constraint$'= 'deny', '$alt\_action$'= 'return CLOSED' ...}
\end{lstlisting}

\item Rule 2:
\label{example2-rule2}
The rule enables REGARD to deny the intermediate AICA-Plan Component action to amend firewall rules. Thus, $RoE_r^2(2)$ is a specific rule that `\textit{denies}' ($r_{c}$(="deny")), requests from `\textit{any source} ($r_{s}$ (="any")) to`\textit{create}' ($r_{v}$(="ADD")) a scope `\textit{10.10.10.10}' ($r_{scope}$(10.10.10.10)) and returns a deny action, `\textit{CLOSED}' ($IRS_{a}$ = $IRS_{D}^2$) as shown in listing ~\ref{list:net-management-system-rule2}.

\begin{lstlisting} [language=YARA, mathescape=true, caption=$RoE_{r}^2(2)$ rule., label=list:net-management-system-rule2]
{ "$r_{id}$"       : "NET-L3-FW",
  "$r_{s}$"        : "any", 
  "$r_{v}$"        : "ADD",
  "$r_{scope}$"    : "10.10.10.10",
  "$r_{c}$"        : "deny",
  "$r_{a}$"        : "return CLOSED"}
\end{lstlisting}

The YARA specification of the $RoE_r^2(2)$ shown in listing \ref{list:net-management-system-rule2} consist of two parts as per the specification in the equation \ref{eq15} and are shown in listings ~\ref{list:net-management-system-rule2-yara1} and ~\ref{list:net-management-system-rule2-yara2} respectively.

\begin{lstlisting} [language=YARA, mathescape=true, caption=YARA Spec for the first part of the \textit{$RoE_r^2(2)$}., label=list:net-management-system-rule2-yara1]
rule NET-L3-FW {
    meta:
        created     = "10/23/2022 09:00:00"
        author      = "ANL"
    strings:
        $\$$source      = "*"
        $\$$int_action  = "ADD" 
        $\$$scope       = "10.10.10.10"
        
    condition: 
        $\$$source and  
            and $\$$int_action and $\$$scope}
\end{lstlisting}

\begin{lstlisting} [language=YARA, mathescape=true, caption=YARA Spec for the second part of the \textit{$RoE_r^2(2)$}., label=list:net-management-system-rule2-yara2]
rule NET-L3-FW {
    meta:
        ...
        '$constraint$'= 'deny', '$alt\_action$'= 'return CLOSED' ...}
\end{lstlisting} 
\end{itemize}

\noindent\textul{Inputs for Case Study 2:}
We take two inputs to showcase the rules pattern matching technique of the rules execution process, as described in Section ~\ref{rules-execution} for a REGARD input, $IRS_{regard_{input}}$. The input from the IRS AICA Plan component contains an intermediate action ($IRS_{ia}$), a source ($IRS_s$), and a target ($IRS_s$) as explained in Section ~\ref{regard-system-design}.

\begin{itemize}
\item Input 1:
\label{example2-input1}
We demonstrate the REGARD execution steps to prevent an AICA-Plan Component plan from allowing a source (`1.2.3.4') to open a TCP SYN connection to the DNS (`10.10.10.20'.) First, the REGARD system receives input from the AICA-Plan component. The input, $IRS_{regard_{input}}^2(1)$, content is shown in the Listing ~\ref{list:web-system-irs-input-1}. The input semantics adheres to the definition shown in the Equation ~\ref{re1}. Then, REGARD applies \{$RoE_{r}^2$\} to the input as mentioned in Equation ~\ref{re2}. Next, it iterates through the rule set and finds a matching rule, $RoE_r^1(1)$. Finally, it overwrites the intermediate action with a CLOSED constraint action, \{$IRS_D^2$\} as $IRS_{regard_{output}}^2(1)$ following the Equation ~\ref{re2} as Listed in ~\ref{list:network-management-system-irs-output-1}.

\begin{lstlisting} [language=YARA, mathescape=true, caption=$IRS_{regard_{input}}^2(1)$ input., label=list:network-management-system-irs-input-1]
{ "$IRS_{ia}$"       : "SYN",
  "$IRS_{s}$"        : "1.2.3.4",   
  "$IRS_{t}$"        : "10.10.10.20" }
\end{lstlisting}

\begin{lstlisting} [language=YARA, mathescape=true, caption=$IRS_{regard_{output}}^2(1)$ output., label=list:network-management-system-irs-output-1]
{ "$action$"       : "Return CLOSED" }
\end{lstlisting}

\item Input 2:
\label{example2-input2}
The REGARD input from the IRS Component plan allows the source (`'1.2.3.4') to add a new firewall rule. However, Rule 2 constrains the intermediate action. The flow is similar to the Input 1 demonstration, but for this input, REGARD finds a matching rule, Rule 2. Listings ~\ref{list:network-management-system-irs-input-2} and ~\ref{list:network-management-system-irs-output-2} show the $IRS_{regard_{input}}^2(2)$ and the $IRS_{regard_{output}}^2(2)$ respectively.

\begin{lstlisting} [language=YARA, mathescape=true, caption=$IRS_{regard_{input}}^2(2)$ input., label=list:network-management-system-irs-input-2]
{ "$IRS_{ia}$"       : "ADD",
  "$IRS_{s}$"        : "1.2.3.4",   
  "$IRS_{t}$"        : "10.10.10.10" }
\end{lstlisting}

\begin{lstlisting} [language=YARA, mathescape=true, caption=$IRS_{regard_{output}}^2(2)$ output., label=list:network-management-system-irs-output-2]
{ "$action$"       : "Return CLOSED" }
\end{lstlisting}
\end{itemize}

\subsection{REGARD RoE Demonstration}
We demonstrate the REGARD system implementation of applying the constraint rules to the intermediate actions by the AICA-Plan components to prevent critical damage to the managed system. For simplicity, we focus on the REGARD RoE execution demonstration and omit the auxiliary subsystem operations. We showcase the second case study implementation of the YARA rules defined in Section ~\ref{example2}. To demonstrate a functional REGARD RoE execution, we implemented the rules definition and their execution logic in one python file. REGARD receives the input from the AICA-Plan component, processes the input, evaluates the processed input with a set of predefined constrain rules, and produces the output for the AICA-Execute components. These steps follow the REGARD system design and its implementation as described in Section ~\ref{regard-system-design}. We list below the RoE python code blocks representing the input, its processing, the rules, their evaluation, the final action list, and the output:

\begin{enumerate}
    \item \textbf{REGARD receive input code block}: REGARD receives an input,$IRS_{regard_{input}}^2(1)$, from the AICA-Plan component. The input is described in Section ~\ref{example2-input1}, more specifically in YARA Specification in listing  ~\ref{list:network-management-system-irs-input-1}. In the python file, the input JSON is presented as a Python string object as shown in listing ~\ref{list:experimental-res-input}.  
\begin{lstlisting} [language=Python, language=Python, caption=$IRS_{regard_{input}}^2(1)$ from AICA- Plan Components., label=list:experimental-res-input]
regard_input = f"""{{ 
"IRS_ia"    : "SYN",
"IRS_s"     : "1.2.3.4",
"IRS_t"     : "10.10.10.20"}}"""
\end{lstlisting}
\item \textbf{Pre-process input code block}: REGARD preprocesses the JSON input as listed in ~\ref{list:experimental-res-input} as a Python function, \textit{preprocess-input}, shown in the code block ~\ref{list:experimental-res-input-transform}. The function transforms the input variable names to match the rules variable names. As an illustration, the function replaces the variable \textbf{$IRS_{ia}$} to match the rules variable names \textbf{int\_action}.
    
\begin{lstlisting} [language=Python, language=Python, caption=$IRS_{regard_{input}}^2(1)$  transformation to RoE input., label=list:experimental-res-input-transform]
def preprocess_input(input: str):
    roe_input = json.loads(input)
    roe_input = input.replace("IRS_ia", "int_action")
    roe_input = roe_input.replace("IRS_s", "source")
    roe_input = roe_input.replace("IRS_t", "scope")
    roe_input = roe_input.replace("1.2.3.4", "*")
    return roe_input
\end{lstlisting}
    \item \textbf{Rule definition code block}:
    REGARD represents the two rules, namely, $RoE_r^2(1)$ and $RoE_r^2(2)$, as a Python string object as shown in listing ~\ref{list:experimental-rule}. These represent the YARA specifications of the two rules in listings~\ref{list:net-management-system-rule1-yara1}, ~\ref{list:net-management-system-rule1-yara2}, ~\ref{list:net-management-system-rule2-yara1}, and ~\ref{list:net-management-system-rule2-yara2}. 

\begin{lstlisting} [language=Python, language=Python, caption=\textrm{Two} $\{RoE_r^2\}$\textrm{ rules.}, label=list:experimental-rule]
roe = f"""
rule NET_L3_DDOS: NET_L3_DDOS {{
    meta:
    created="10/23/202209:00:00"
    author="ANL"
    constraint="deny"
    alt_action="return CLOSED"

    strings:
    $source="*"
    $int_action="SYN"
    $scope="10.10.10.20"

    condition:
    $source
    and $int_action and $scope}}
rule NET_L3_FW: NET_L3_FW {{
    meta:
    created="10/23/202209:00:00"
    author="ANL"

    strings:
    $source="*"
    $int_action="ADD"
    $scope="10.10.10.20"

    condition:
    $source
    and $int_action and $scope}}"""
\end{lstlisting}

    \item \textbf{Rule evaluation code block}: REGARD implement this block as two python statements, namely \textit{compile} and \textit{match}, adhering to the YARA python library semantics. The first statement uses \textit{yara.compile} to compile the two rules, defined as Python strings in listing ~\ref{list:experimental-rule}. The second statement uses \textit{match} to the processed input, listed in ~\ref{list:experimental-res-input-transform} with the two compiled rules. We list the compile and match steps in the code block listing ~\ref{list:experimental-res-roe-engine}.

\begin{lstlisting} [language=Python, language=Python, caption=RoE Engine compilation and evaluation., label=list:experimental-res-roe-engine]
def roe_engine (lh: str):
    compiled_roe = yara.compile(source=roe)
    matches = compiled_roe.match(data=lh)
\end{lstlisting}

    \item \label{final-action-demo} \textbf{Final action creation code block}: 
    REGARD uses an \textit{if-else} structure as in listing ~\ref{list:experimental-res-roe-engine-match} to determine either to constrain the AICA-Plan intermediate action. The \textit{if} block creates the final evaluated action $\{IRS_a\}$ set ($IRS_p$), per RoE engine evaluation code block listing ~\ref{list:experimental-res-roe-engine}, for the input listing ~\ref{list:experimental-res-input}, using the rules in listing ~\ref{list:experimental-rule}. The \textit{else} block implements the logic if there is no matching rule for the input. The else block also implements a default rule matching case that denies the intermediate action following the \textit{allow list} approach as defined in the rules definition description (See Section ~\ref{dnd}). The else block creates the deny action, $IRS_{D}^2$, as defined in equation ~\ref{eq17}.
\begin{lstlisting} [language=Python, language=Python, caption=RES action set creation., label=list:experimental-res-roe-engine-match]
    regard_output = {}    
    if matches:
        actions = list()
        for match in matches['main']:
            rh = match['meta']
            rh_constraint = rh['constraint']
            rh_alt_action = rh['alt_action']
            if rh_constraint == 'deny':
                actions.append(rh_alt_action)
        regard_output['actions'] = actions
        regard_output['input'] = original_input
    else:
        # default is deny all intermediate actions
        regard_output['actions'] = ['CLOSED']
        regard_output['input'] = original_input
    return json.dumps(regard_output, indent=2)
\end{lstlisting}
    \item \textbf{IRS REGARD output creation code block}: REGARD creates a JSON formatted output as in listing ~\ref{list:experimental-res-output}. The output, $IRS_{regard_{output}}^2(1)$,  contains two items: a list of final actions, $\{IRS_a\}$, and the initial input, $IRS_{regard_{input}}^2(1)$, to the system. The action list is constructed from the final actions created in the code block listing ~\ref{list:experimental-res-roe-engine-match}. The initial input is the same information the REGARD system received in the code listing ~\ref{list:experimental-res-input}. 

\begin{lstlisting} [language=Python, language=Python, caption=$IRS_{regard_{output}}^2(1)$ to AICA IRS Execute Components., label=list:experimental-res-output]
{"actions": [
    "CLOSED"
  ],
  "input": "{ \"IRS_ia\"    : \"SYN\",\"IRS_s\"     : \"1.2.3.4\",\"IRS_t\"     : \"10.10.10.20\"}"}
\end{lstlisting}
\end{enumerate}

In this section we demonstrated the flexibility of the RoE rule definition that an organizational domain expert uses to constrain the AICA-Plan Component actions. Without the REGARD Component, some AICA-Plan Component actions can cause irreversible detrimental damage to the managed systems. Thus, REGARD acts as a security gatekeeper checking every AICA-Component action based on the constrained rules and either denying it or letting it pass through. Hence, REGARD boosts the robustness of AICA SA-ACS IRS by adding another layer of validation before taking any action on the managed systems.


\section{Conclusion}
\label{conclusion}

In this paper, we describe Rules of EngaGement for Automated cybeR Defense (REGARD) system architecture, design, roles and responsibilities, and rule definitions. The REGARD system extends the Self-Adaptive Autonomic Computing System (SA-ACS) paradigm and is based on the Automated Intelligent Cyber defense Agents (AICA) architecture.  REGARD empowers the security administrators to constrain the intermediate actions from the AICA Plan Components while mitigating a security threat in an automated fashion. REGARD holds a set of Rules of Engagement (RoE) to protect the managed system according to the instructions provided by the domain expert. These rules help limit the action of the IRS on the managed system in compliance with the recommendations of the domain expert.

In REGARD, we pursue a template approach to define and manage the rules. In addition, we leverage the Regular Expression string-matching pattern to evaluate the rules for the intermediate actions. We provide details of REGARD execution, management, operation, and conflict resolution for Rules of
Engagement (RoE) to constrain the actions of an automated IRS. This includes a description of the rules engine and a specific software implementation as a prototype. REGARD follows a micro-service architecture with two components aiming to scale and secure the individual components. 

We also present REGARD system implementation, security case studies for cyber defense, and RoE demonstrations. We showcase the execution of REGARD on two managed systems, namely, a web system and a network infrastructure management system. 

In the future, we plan to more closely integrate the REGARD system with the AICA system, utilize Graph Neural Network to evaluate the rules, and enhance the robustness of the conflict resolution part of the rules engine.

\section*{Acknowledgement}
This work was supported by the INSuRE (Information Security Research and Education) Project in collaboration with Argonne National Laboratory, and National Science Foundation grants (\#1565484, \#2133190). The views and conclusions are those of the authors.
\bibliographystyle{unsrt}
\bibliography{regard.bib}


\end{document}